# DIFFERENTIAL STROBOSCOPE FOR PHYSICS LAB

**RAJU BADDI**
National Center for Radio Astrophysics, TIFR, Ganeshkhind P.O. Bag 3, Pune University Campus, PUNE 411007, Maharashtra, INDIA; baddi@ncra.tifr.res.in

## ABSTRACT

Conventional stroboscope uses flashes of light to make a rapidly moving object visible. This is achieved by throwing repetative pulses of white light on the object of specific frequency. Here an alternative approach is given in which two different colors of light(●/●) are used to make a rapidly moving object visible. As such this method has two light pulses of different colors within a single strobe period. Such a stroboscope has its own advantages in experiments over the conventional one. This article describes the construction of this stroboscope in all detail. Various aspects of its performance are discussed pictorially with reference to a simple rotating disc. This stroboscope can be used in the study/quantitative-assessment of rapidly moving objects which by default are taken to be cyclic in nature.

## I. INTRODUCTION

It is difficult for the naked eye to make out something meaningful about a rapidly moving object like for instance a rotating ceiling fan. It is almost impossible for a person unaided by any tools to say something about the number of blades or the shape of these blades. However by a wink of the eye one can see the rotating fan for a short duration of time. By repetative rapid winking one can see the blades of the fan become visible and so does their shape. This rapid winking presents snapshots of the moving object for short intervals of time to the unaided eye. During these snapshots the moving object does not have enough time for appreciable displacement. As such these snapshots capture a static picture of the object in question. Hence the eye and the human brain have a more meaningful assessment of the presentation.

From the above discussion it is clear that intermitent illumination of moving objects would also present rapid snapshots. This intermittent illumination can be carried out at a much higher rate than that possible for the winking of the human eye. Hence intermittent illumination and its understanding forms a study of rapidly moving objects. This approach has been called as stroboscopy. Stroboscopic method to study rapidly moving objects has been in practice for more than 100 years. By stroboscopic method one can see details of rapidly moving objects and assess their displacement rate(Chadwick 1939). For instance by throwing stroboscopic light it is possible to deduce the rotation rate of a disc without making any physical connection with it. When the frequency of rotation of the disc becomes equal to the frequency at which light pulses are thrown on the disc one sees a static picture of the disc as shown in Figure 1(D). The frequency of the light pulses can be independently measured which now would be equal to that of the rotation of the disc. However such a static picture is again possible when the strobe frequency is half of the rotation frequency as in Figure 1E. Except for a slight broadening(assuming duty cycle remains the same) in the illuminated circular patch. So it is difficult to distinguish between states D and E in Figure 1. At this point we consider some of the aspects of stroboscopic light. Which can be represented on a graph sheet as shown in Figure 2. The repetative period **T** of the stroboscope consists of a light period($\mathbf{T_{on}}$) and a blank period($\mathbf{T_{off}}$) with $\mathbf{T = T_{on} + T_{off}}$. $\mathbf{T_{on}}$ is the period for which the moving object is illuminated. It is normally desired to keep $\mathbf{T_{on}}$ brief so that the object does not suffer appreciable displacement during $\mathbf{T_{on}}$ yet it sufficiently illuminates the object. This criterion is gathered in the duty cycle **ξ** of the light train, $\boldsymbol{\xi = T_{on}/(T_{on}+T_{off})}$. A low value of **ξ** ensures sharp picture of the object but also means low illumination. As one seeks sharper picture by reducing **ξ** one looses upon the brightness of the object for a given peak illumination. The value of **ξ** is normally set on trial and error basis by careful observation during the experiment.

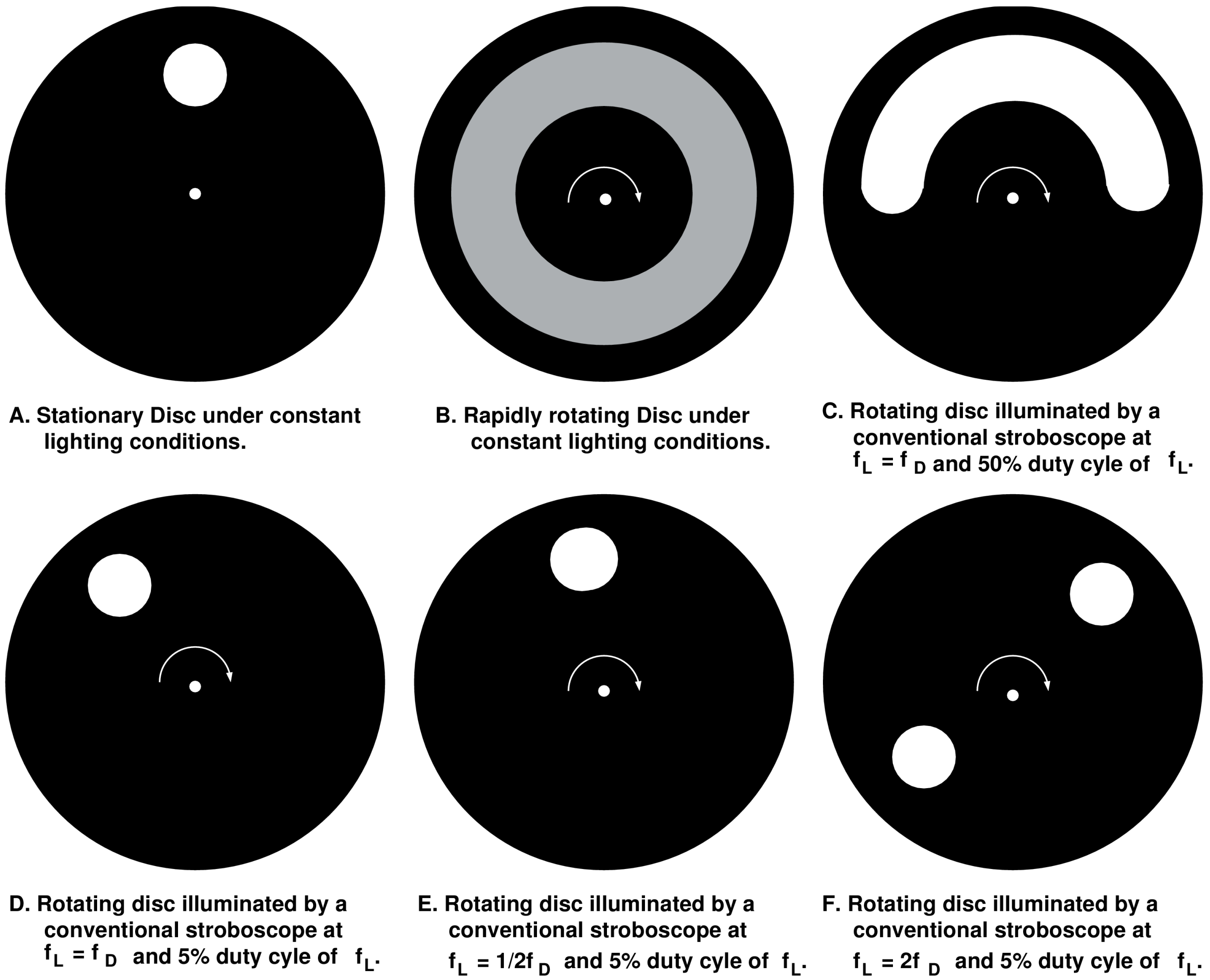

*Fig 1: A black disk with a circular white patch stationary/rotating under ordinary constant lighting condition and conventional stroboscopic illumination(white). A few cases of stroboscopic frequency and duty cycle are shown. Duty cycle of the stroboscope( $f_L$) is the fraction of the repetative period for which the light appears. Figure 3 shows the same disk under differential stroboscope for comparison.*

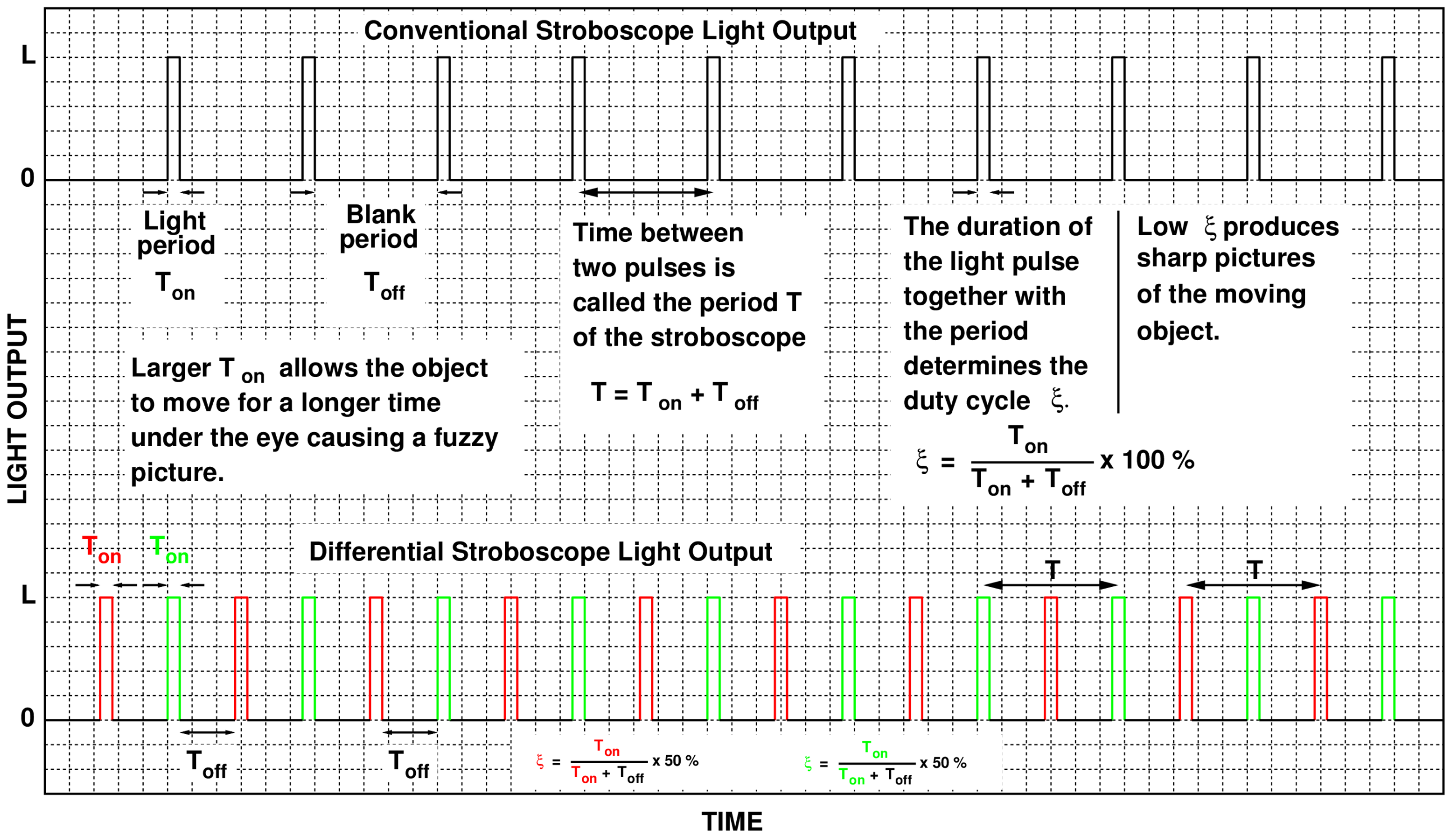

*Fig 2: Graphical representation of light pulses from a stroboscope. Different parameters relevant to the light output are shown in the plot. The top shows the light pulses from the conventional strobe while the lower train shows the differential strobe. Lower panel: Same symbols mean equal values.*

Conventional stroboscopes employ either gas discharge lamps(typically xenon/neon) or LEDs and throw light of one color. Here use of two colors of light has been made. The light output pulse train in the proposed method has been shown in the lower panel of Figure 2. The results of observations on a rotating disc are displayed in Figure 3 and can be directly compared with the corresponding counter parts in Figure 2. One of the immediate advantages that is apparent is recognition of either the object rotating at a multiple frequency of the strobe(Figure 3E) or the strobe throwing light at a multiple frequency of the rotating object(Figure 3F). These kind of situations are easy to make out under differential strobe rather than under conventional strobe. The same principle applies to any other cyclic process like a vibrating reed or the flapping of wings of an insect(Chadwick 1939) or perhaps the movement of insect legs. It should be noted that the duty cycle w.r.t light of one color remains the same for differential strobe. As light of one color is emitted only once every period. So the duty cycle for each color is just the same as for conventional strobe. The next section describes the constructional details of the Differential Stroboscope.

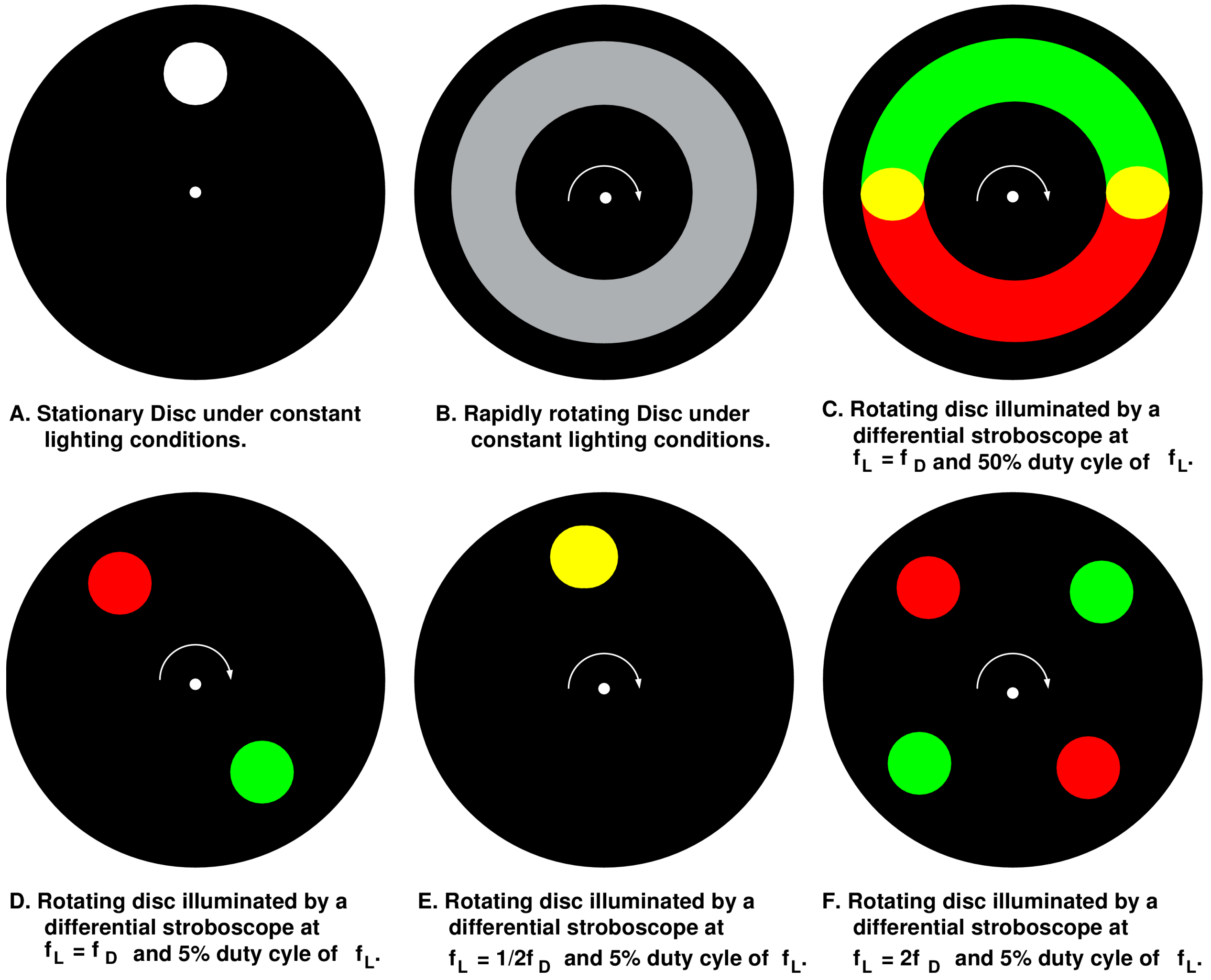

*Fig 3: A black disk with a circular white patch stationary/rotating under ordinary constant lighting condition and differential stroboscopic illumination(red/green). It can be seen that differential stroboscope has the advantage of distinguishing cases where the rotation frequency is multiple/fractional times the strobe frequency. Cases D and E are good examples. In the case of conventional strobing except for the slight broadening it is difficult to distinguish between these two cases as can be seen from Figure 1. However differential strobing presents them completely different from eachother.*

## 2. THE DIFFERENTIAL STROBOSCOPE

The differential stroboscope is similar to the conventional stroboscope except that it uses light pulses of two colors. These pulses are emitted within a single period of the stroboscope. So for a given period if the conventional strobe throws a single pulse of monochrome light. The differential strobe throws two pulses of light of different colors (●/●) separated by half of the period. The various outcomes when a rotating disc with a single white patch is differetially strobed are given in Figure 3. It can be seen from these figures that the immediate harmonics can be easily ruled out(Figure 3D,E,F). Light pulses of different colors can be easily obtained using light emitting diode(LED). The ubiquitous timer chip NE555(6) can be used to generate the required repetative pulse frequency and the pulse width for the strobe light. The NE555 can be used either as reliable astable multivibrator or as a one-shot pulse generator(commonly called *monostable* operation). By using 555 in both these modes and connecting them in series it is possible to simultaneously control both the frequency as well as the light pulse width. The package NE556 provides dual 555 timer chip and hence is suitable for our purpose. The astable mode is used to repetatively trigger the monostable mode causing production of voltage pulses of specific widths from the monostable. Both the frequency of the astable and the pulse width of the monostable can be controlled independently using simple variable resistors(VR1/VR2) and a set of fixed capacitors. Where as the astable gives control of the frequency of the strobe light the monostable provides the control over the strobe duty cycle. Further the frequency of the astable can be precisely measured using inexpensive CMOS circuitry employing simple gates, counters, flipflops and a common quartz clock timing chip. The frequency meter immediately reads the frequency of the strobe and hence is reliable than depending on dials or calibration. Further it may also be desirable to have provision for changing the intensity of strobe light or blanking of one of the colors to immediately have the conventional stroboscope. All these are taken care of in the proposed design whose schematic circuit diagram is shown in Figure 4. Together with the above discussed provisions with this design it is also possible to shift the position of stationary patches to a desired orientation using the switches called *strobe phase delay* and *strobe phase advance*. The strobe can be operated in either Continuous Mode(C) or Pulsed Mode(P) as provided by switch S0. The continous mode is essentially a direct jump to 50% duty cycle where as the pulsed mode is for lower duty cycle operation controlled through VR2. The frequency is read on Binary Coded Decimal(BCD) counters, CD4518. A common quartz clock chip(example: UM3252) together with logic gates(CD4011) and flipflops(CD4013) is used to generate a gate window of one second for G on momentary depression of a switch(FRQ) during which oscillations from the astable are passed on to the appropriate counter(CD4518) trigger input. This arrangement forms the frequency meter of the stroboscope which can read upto a count of 9999 or approximately a frequency of 10kHz. NE556A forms the astable which triggers the NE556B monostable through a network of inverter gates(CD4069) and RC circuits. The monostable is triggered both on rising as well as falling edge of NE556A output. The polarity of output of the astable(which here has been arranged to produce square waves with nearly 50% duty cycle) also controls which color of the LED is fired on the generation of the pulse from the monostable. This is acheived by driving the base of the transistors Q1 and Q2 through inverter gates which are directly coupled to the astable. Each transistor(Q1, Q2) in turn drives a set of 4 LEDs of one color(●/●). Transistor Q3 drives the other end of these LEDs and is driven by the pulse from the monostable for a specific period as desired by setting VR2. Q4 blanks the frequency display LEDs during the time of measurement for reliable counting. A momentary depression of pushbutton FRQ is required for every fresh measurement of frequency. The 1-second pulses from the quartz clock pulser are obtained from the two outputs of UM3252 which drive the electromagnetic coil. They are applied to a comparator LM393 whose output produces a positive going pulse every second. This is indicated by the flashing of a red LED. FRQ is preferrably depressed when LED1 goes off.

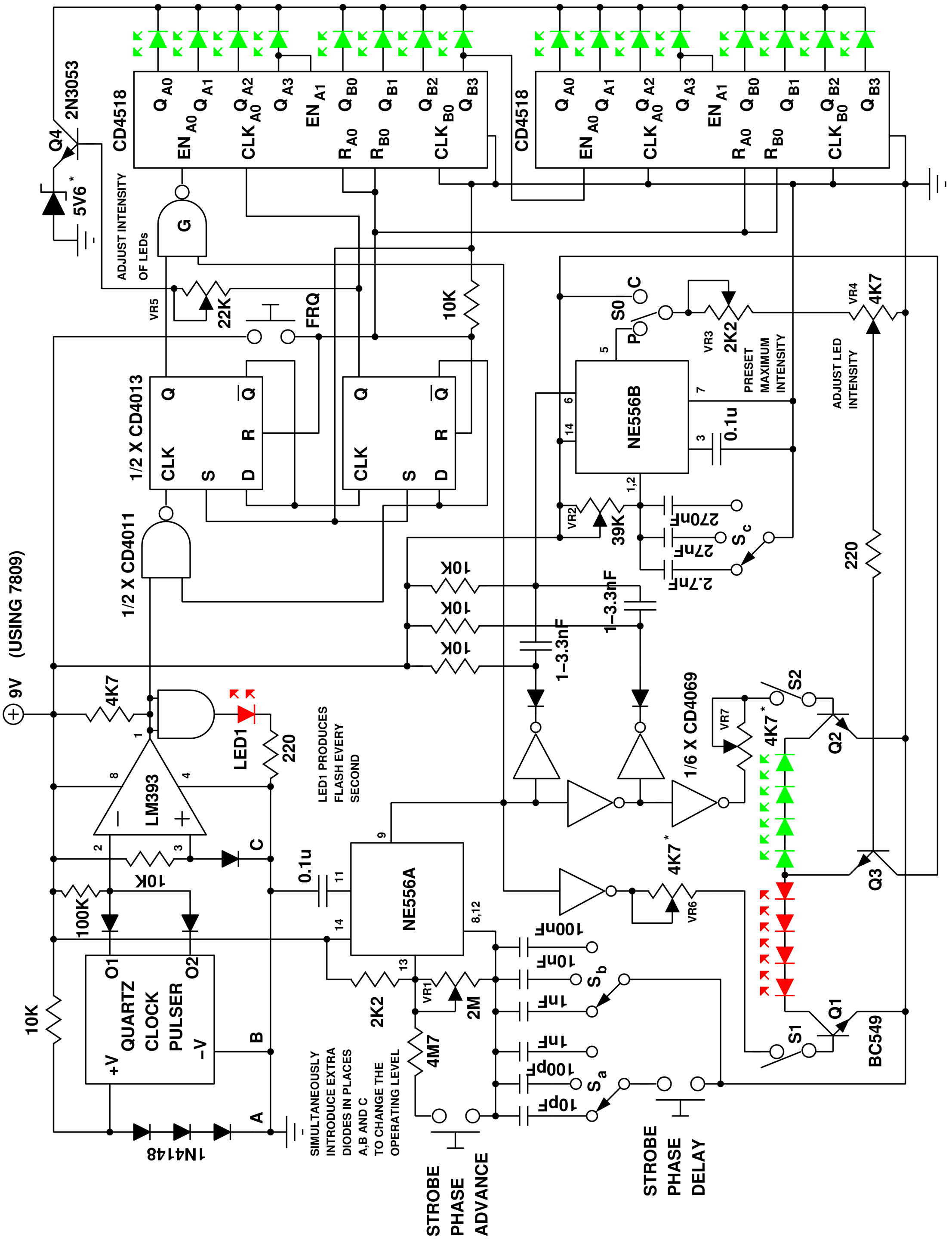

*Fig 4: Proposed schematic circuit diagram of Differential Stroboscope. The values with a star have to be appropriately selected and can vary from what has been quoted. Read the text for further description. The range goes as 3.6 – 50Hz,36 - 500Hz and 360 – 5kHz from right of Sa,Sb or Sc which have a common handle.*

Figure 5 shows a diagram giving the layout of the circuit on a general purpose board. The proposed arrangement for the apparatus is shown as an illustration in Figure 6. Figure 7 shows the operations of strobe phase advance/delay pictorially. Figure 8 shows the complete post having the strobe LEDs.

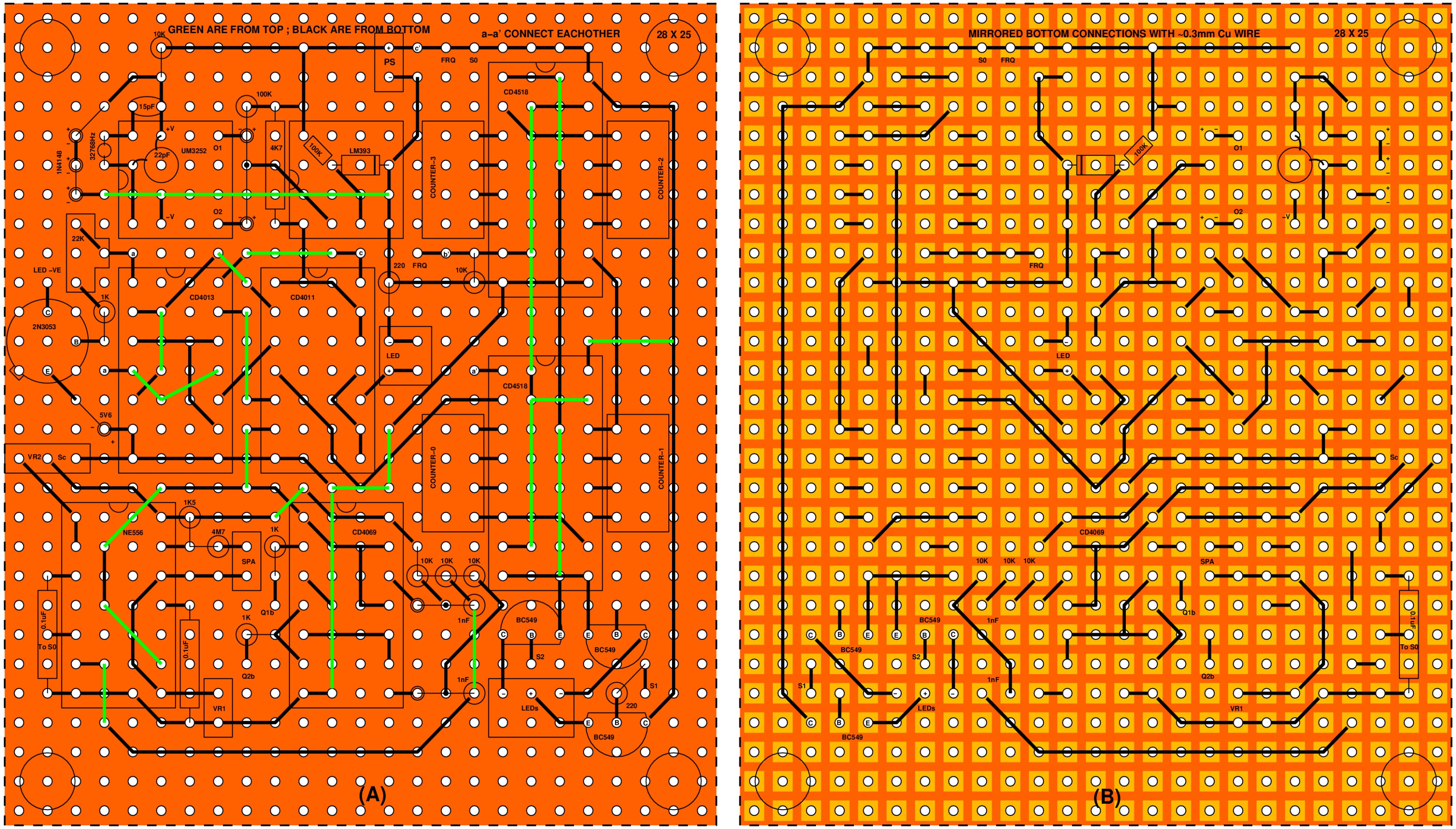

*Fig 5 : Circuit layout of Differential Stroboscope on a General purpose board(28 x 25 holes). The green connections are from top while the black are from bottom. The left figure gives a X-ray view of the layout and component placement while the right gives the mirrored bottom connections.*

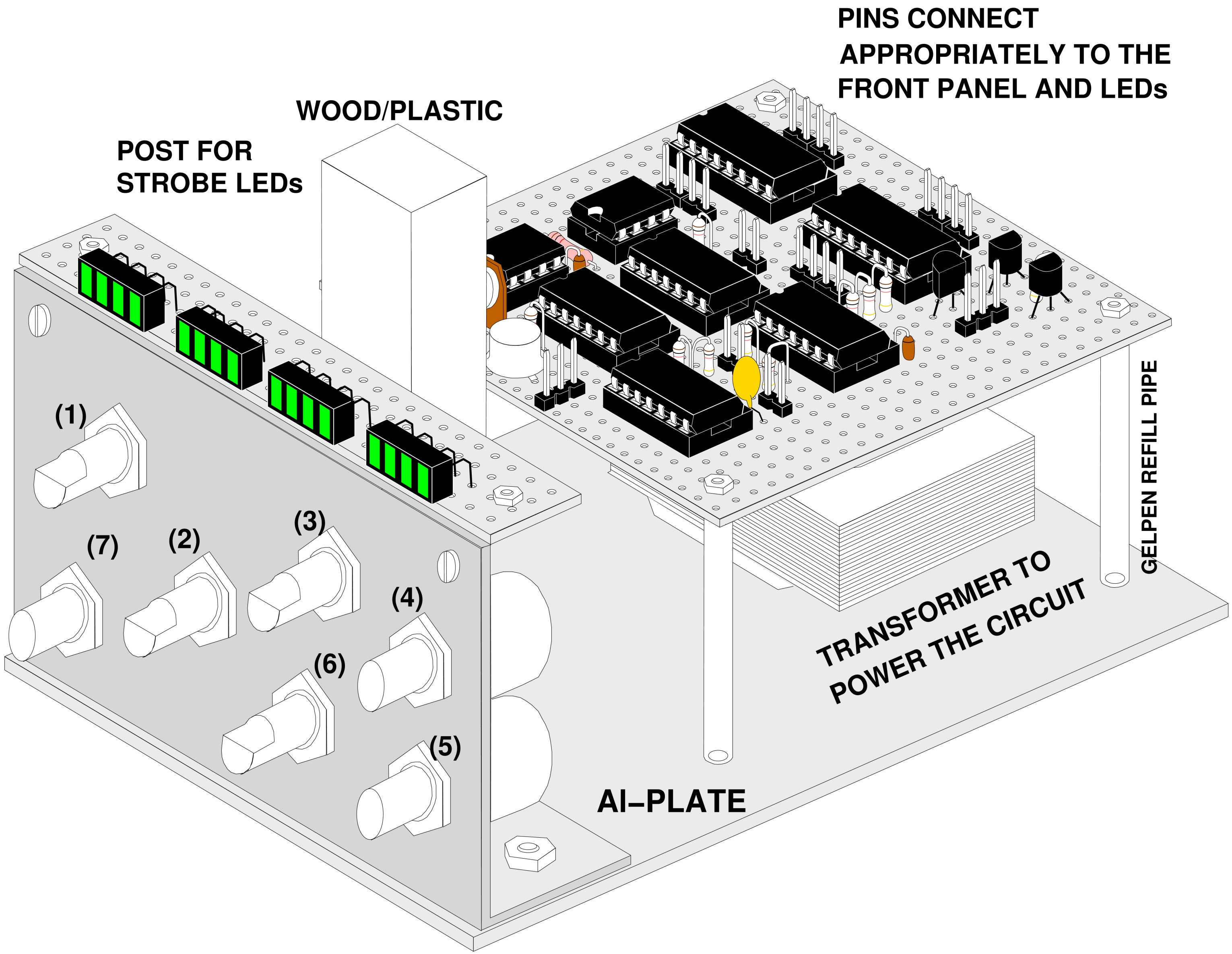

*Fig 6 : Indicative illustration showing how the Stroboscope could be built. (1)-Frequency adjust (2)-Frequency range (3)-Pulse width (4)-Strobe Advance (5)-Strobe Delay (6)-Pulse intentisty (7)-Frequency sample. Further refer Figure 8 for details regarding fixing strobe LEDs to the post.*

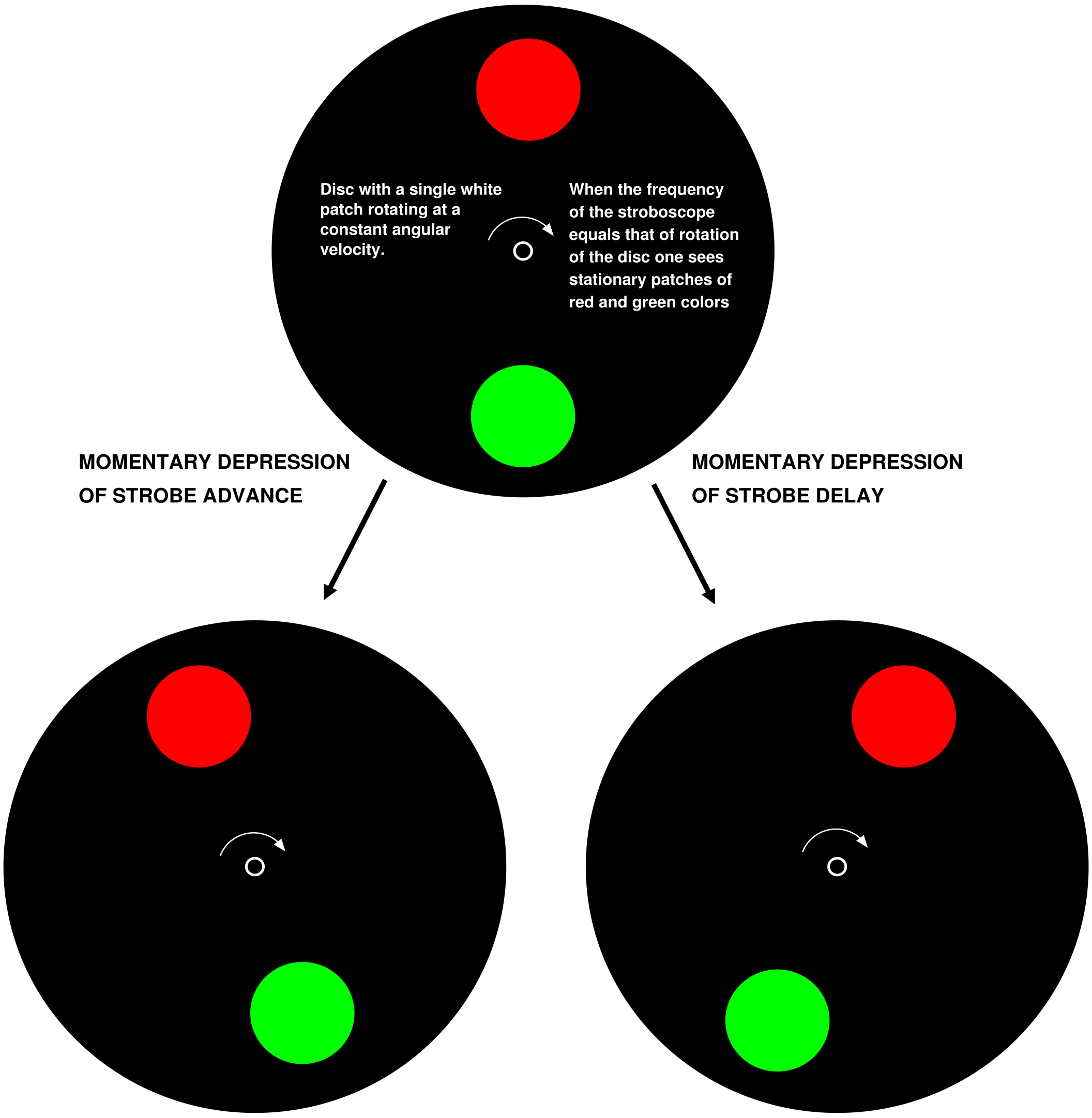

*Fig 7 : Operation of Strobe Phase Advance and Strobe Phase Delay switches.*

## 3. IMPORTANT NOTES REGARDING THE STROBE/SCHEMATICS

1. The values of the components with a star has to be selected appropriately by trial and error according to ones requirement. For instance the variable resistors VR6 and VR7 can be choosen to be higher, say 47K instead of 4K7. With these it would be possible to adjust individual brightness of LEDs of two colors. However in the front panel of Figure 6 they have not been accomodated. Interested reader can do so. The same is true for 5V6 zener if desired it can be reduced to 5V1 or 4V8 for more brighter LED glow.
2. The operating level of LM393 can be varied by introducing extra diode(s)(1N4148 or a zener say 3.0V) in series common to A, B and C as shown in Figure 4. It should be noted that the quiescent reference for LM393 is provided by the voltage drop across a single silicon diode which would be just ~0.6V above the ground voltage as of now.
3. Strobe Phase Advance/Delay provide an easy way to change the stationary position without disturbing the frequency setting.
4. Further the blanking switches(S1 & S2) are not accomodated on the front panel. These would be especially useful if one needs to quickly configure the strobe as monochrome. By having white LEDs instead of one set of color it is also possible to have the conventional strobe without any change of hardware. However one will have to blank the other color.
5. VR3 is used to set an upper limit on the light intensity while VR4 provides intensity control.
6. NE556A does not produce an exactly 50% duty cycle waveform due to finite value of resistance between pin 14 and 13. However if desired a lower value than 2K2 can be used.
7. If desired the supply voltage can be increased to accomodate more/brighter LEDs.

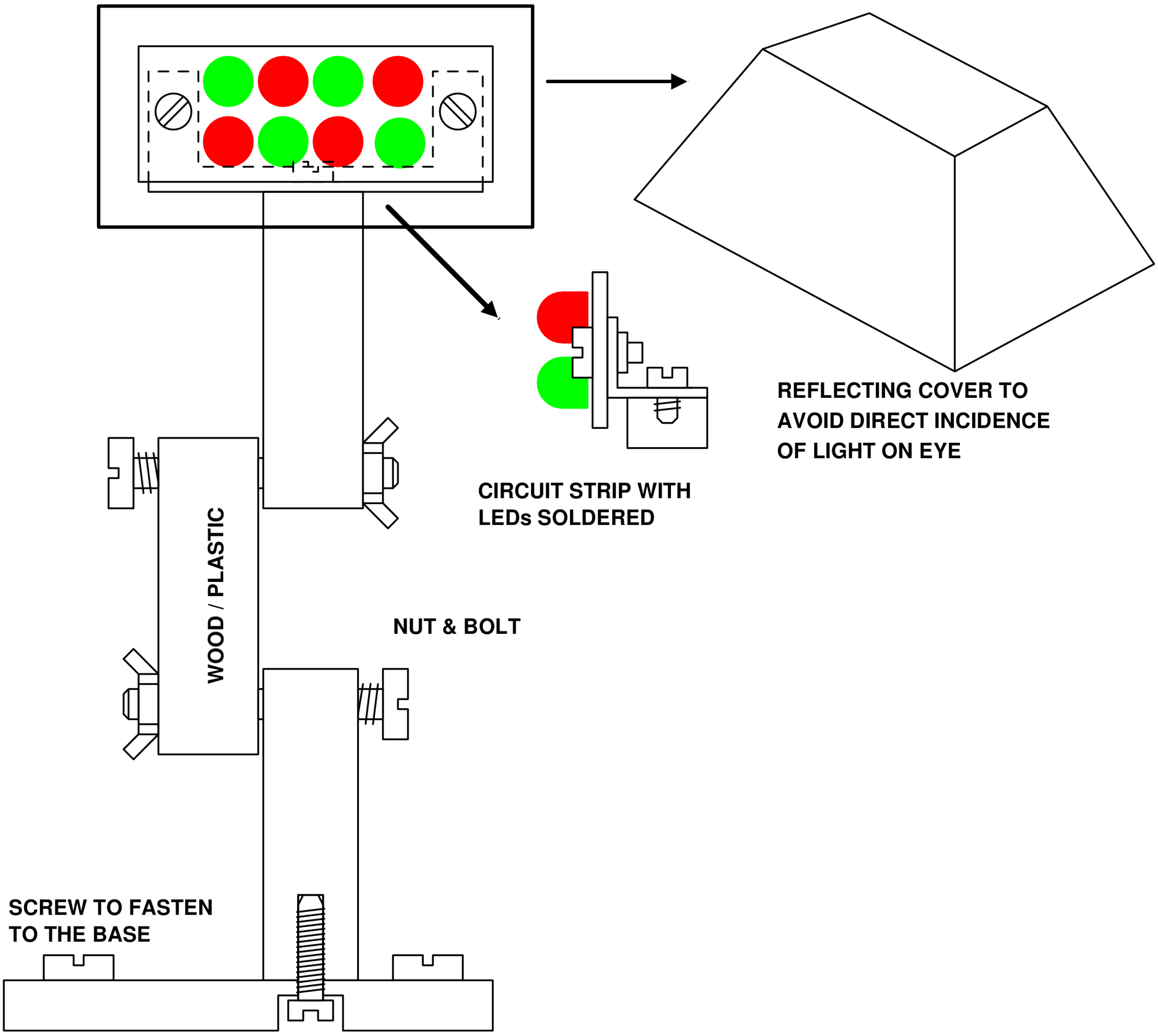

*Fig 8 : The complete post shown partially in Figure 6 which connects to the aluminium base plate. Figure 6 shows only the first stem.*